\documentclass[conference]{IEEEtran}

\usepackage{svg}
\usepackage{eurosym}
\usepackage[cmex10]{amsmath}
\usepackage{multirow}
\usepackage{url}
\usepackage[normalem]{ulem}

\newenvironment{mydescription}[1]
  {\begin{list}{}%
   {\renewcommand\makelabel[1]{##1 \hfill}%
   \settowidth\labelwidth{\makelabel{#1}}%
   \setlength\leftmargin{\labelwidth}
   \addtolength\leftmargin{\labelsep}}}
  {\end{list}}

\let\savehline\hline
\def\sphline{\noalign{\vskip3pt}\savehline\noalign{\vskip3pt}}

\IEEEoverridecommandlockouts
\usepackage{fancyhdr}
\usepackage{cite}
\usepackage{amssymb,amsfonts}
\usepackage{algorithmic}
\usepackage{graphicx}
\usepackage{textcomp}
\usepackage{xcolor}
\usepackage{array}    
\usepackage{fancyhdr}
\def\BibTeX{{\rm B\kern-.05em{\sc i\kern-.025em b}\kern-.08em
    T\kern-.1667em\lower.7ex\hbox{E}\kern-.125emX}}

\usepackage{titlesec}

\usepackage{amsmath}

\renewcommand{\thesection}{}  
\renewcommand{\thesubsection}{} 
\renewcommand{\thesubsubsection}{}

\titleformat{\section}[block]{\bfseries\normalsize}{\thesection}{0pt}{}  
\titleformat{\subsection}[block]{\itshape\normalsize}{\thesubsection}{0pt}{}  
\titleformat{\subsubsection}[block]{\itshape\normalsize}{\thesubsubsection}{0pt}{} 

\begin{document}
\rmfamily

\title{{\fontsize{16}{19}\selectfont \textbf{Electric power system security: the case for an integrated cyber-physical risk management framework}}}


\author{

    \IEEEauthorblockN{{\fontsize{12}{14}\selectfont Efthymios Karangelos}}
    \IEEEauthorblockA{\textit{\fontsize{10}{12}\selectfont School of Electrical and Electronic Engineering} \\
    \textit{\fontsize{10}{12}\selectfont University College Dublin}\\
    \fontsize{10}{12}\selectfont Dublin, Ireland\\
    \fontsize{10}{12}\selectfont efthymios.karangelos@ucd.ie}
    \and
   
    \IEEEauthorblockN{{\fontsize{12}{14}\selectfont Louis Wehenkel}}
    \IEEEauthorblockA{\textit{\fontsize{10}{12}\selectfont Department of Electrical Engineering and Computer Science} \\
    \textit{\fontsize{10}{12}\selectfont Universit\'{e} de Li\`{e}ge}\\
    \fontsize{10}{12}\selectfont Li\`{e}ge, Belgium \\
    \fontsize{10}{12}\selectfont l.wehenkel@uliege.be}
    }
\maketitle
\thispagestyle{fancy}

\begin{abstract}
This paper concerns the security of the electric power transmission grid facing the threat of malicious cyber-physical attackers. We posit that there is no such thing as perfectly effective cyber-security. Rather, any cyber-security measure comes with the possibility
that a highly skilled attacker could (eventually find a way to) bypass it. On these grounds, we formulate a tri-level decision making problem seeking to co-optimize preventive physical and cyber-security measures under uncertainty on the ability of an exogenous cyber-physical attacker to  overcome the latter. Preventive physical security measures refer to the \emph{ex-ante} procurement of reserve capacity,
 which translates into 
 ramping restrictions in real-time. Cyber-security measures refer to updating the  firewall rules so as to 
 impede an intruder from taking over the cyber infrastructure of the grid and disconnecting power generators and transmission branches.  We adopt standard  assumptions to formalize  the inner optimization problems corresponding to the cyber-physical attacker and power grid operator and focus on uncertainty management 
at the uppermost level of the problem. Our findings establish that physical- and cyber-security measures are non-exchangeable complements in keeping the power grid operation secure.
\end{abstract}

\begin{IEEEkeywords}
 cyber-physical systems, electric power system  risk management.
\end{IEEEkeywords}

\section{Notation}

The main symbols used in this text are defined as follows.
\subsection{Sets}

\begin{mydescription}{$up_{g,i}$}
              \item[$\mathcal{A}$]{Cyber-physical attackers with advanced capabilities (\textit{i.e.}, able to by-pass cyber-security measures);}
        \item[$\mathcal{B}$]{cyber-physical attackers with basic capabilities (\textit{i.e.}, unable to by-pass cyber-security measures);}  
    \item[$\mathcal{G}$]{generating units;}
    \item[$\mathcal{N}$]{buses;}
	\item[$\mathcal{T}$]{transmission branches.}
 \end{mydescription}

\subsection{Parameters}
\begin{mydescription}{$up_{g,i}$}
    \item[${C}^{D}_g$]{Dispatch cost of generator $g \in \mathcal{G}$ -- in monetary units/MWh (mu/MWh);}
    \item[${C}^{R}_g$]{reserve capacity cost of generator $g \in \mathcal{G}$ -- in mu/MW;} 
    \item[${C}^{RD}_g$]{reserve deployment (\textit{i.e.}, redispatch) cost of generator $g \in \mathcal{G}$ -- in mu/MWh;} 
    \item[$C^{L}$]{value of lost load -- in mu/MWh;}
    \item[${C}^{Z}$]{cost of updating firewall rules  -- in mu/device;}
    \item[$D_n$]{active power demand at bus $n \in \mathcal{N}$;}
    \item[$F_{t}$]{capacity of transmission branch $t \in \mathcal{T}$;}
    \item[${P}_g$]{active power generation capacity of generator $g \in \mathcal{G}$;} 
        \item[$W_{i}$]{resource limit of cyber-physical attacker $i \in \mathcal{A}\cup\mathcal{B}$;}
    \item[$X_{t}$]{reactance of transmission branch $t \in \mathcal{T}$;}
     \item[$Z$]{resource limit of cyber-physical planner;}
	\item[$\gamma_{g,n}$]{auxiliary  parameter, taking the value of 1 to indicate the bus at which generator $g \in \mathcal{G}$ is connected;}
  \item[$\pi_{i}$]{manifestation probability of cyber-physical attacker $i \in \mathcal{A}\cup\mathcal{B}$;}
    \item[$\tau_{t,n}$]{auxiliary parameter taking  the value of 1(-1) to indicate the sending(receiving) buses of branch $t \in \mathcal{T}$.}  
  \end{mydescription}
 
\subsection{Variables}
 
\begin{mydescription}{$up_{g,i}$}
    	\item[$p_g$]{Active power dispatch of generator $g \in \mathcal{G}$ in the pre-attack state --  by the planner;}
     	\item[$r_g$]{procured reserve capacity from generator $g \in \mathcal{G}$ --  by the planner;}
            \item[$z_n$]{binary variable, taking the value of  1 to indicate the update of the firewall rules at bus $n \in \mathcal{N}$ --   by the planner;}
              \item[$w_{n,i}$]{binary variable, taking the value of 0 to indicate successful              intrusion into the cyber sub-system of bus $n \in \mathcal{N}$ -- by  cyber-physical attacker $i \in \mathcal{A}\cup\mathcal{B}$;} 
            \item[$v_{t,i}$]{binary variable, taking the value of 0 to indicate the disconnection of transmission branch $t \in \mathcal{T}$ -- by  cyber-physical attacker $i \in \mathcal{A}\cup\mathcal{B}$;} 
            \item[$y_{g,i}$]{binary variable, taking the value of 0 to indicate the disconnection of generator
            $g \in \mathcal{G}$ -- by  cyber-physical attacker $i \in \mathcal{A}\cup\mathcal{B}$;} 
            \item[$up_{g,i}$]{upward active power redispatch of generator $g \in \mathcal{G}$ in the post-attack state corresponding to attacker
              $i \in \mathcal{A}\cup\mathcal{B}$ -- by the power grid operator;}
            \item[$dp_{g,i}$]{downward active power redispatch of generator $g \in \mathcal{G}$ in the post-attack state corresponding to attacker $i \in \mathcal{A}\cup\mathcal{B}$ -- by the power grid operator;}
            \item[$l_{n,i}$]{involuntary load shedding at node $n \in \mathcal{N}$  in the post-attack state corresponding to attacker  $i \in \mathcal{A}\cup\mathcal{B}$ -- by the power grid operator;}
	\item[$f_{t,i}$]{active power flow  through branch $t \in \mathcal{T}$  in the post-attack state corresponding to attacker  $i \in \mathcal{A}\cup\mathcal{B}$;}
	\item[$\theta_{n,i}$]{voltage angle at node $n \in \mathcal{N}$  in the post-attack state corresponding to attacker $i \in \mathcal{A}\cup\mathcal{B}$.}
 
\end{mydescription}

\section{Introduction}\label{intro}

Advanced digital technology is vital for the functioning of today's electric power system. Indeed, the modern power grid relies on its cyber sub-system for most monitoring, protection, control and optimization functionalities and operates as a complex  \emph{cyber-physical} system. This essential technology however may also pose a severe vulnerability. The risk that a malicious adversary may intentionally exploit the power grid's cyber sub-system so as to disrupt electricity supply has unfortunately been validated in practice \cite{sullivan2017cyber}. As a result electric power system cyber-security is nowadays recognized as critical and there is a growing development of standards, rules and regulations \cite{ALVAREZALVARADO2024109149}. 

\subsection{Related literature}

The question of securing the grid operation while facing the threat of an intelligent cyber-physical  attacker is actively pursued by the research community. Most research works rely on multi-level optimization to explicitly model the interrelated decisions of malicious cyber-physical attackers and power grid so-called ``defenders'' (\textit{i.e.}, planners and operators) \cite{karamdel2022}. Bi-level optimization problems are cast from the  perspective of a cyber-physical attacker seeking to optimize her attack vector while explicitly accounting for the optimal reaction of the power grid operator. Such models can be used in an assessment context, so as to identify which vulnerabilities a cyber-physical attacker could target to exploit. Tri-level problems are cast from the (preceding) perspective of a preventive planner. The goal here is to identify preventive measures to counteract a cyber-physical attacker's optimal strategy, in turn  maximizing her own objective while anticipating the grid operator's reaction.  He \textit{et al.} \cite{HE2021106903} consider the problem of facing a cyber-physical attacker capable of both injecting false data and physically disconnecting transmission lines. An innovative feature of this work is the allocation of Distributed Generators (DGs) in the post-attack stage, translating mathematically into binary variables at the lowest-level optimization problem modeling power system operation.  The considered preventive measures only act on the physical infrastructure of the grid by hardening transmission components. In contrast, Lai \textit{et al.} \cite{LAI2019204} focus on choosing cyber-security preventive measures only. These authors address the problem of placing firewalls so as to block unauthorized access to protection relays that could be exploited by a cyber-physical attacker seeking to cause a cascading failure. Qin \textit{et al.} \cite{QIN2023120961} also consider the  placement of firewalls to impede unauthorised access to the grid's relays. The novel feature of this work is the consideration of  detailed substation topologies. Both groups of authors in \cite{LAI2019204} and \cite{QIN2023120961} adopt  the assumption that firewalls are perfectly effective countermeasures, blocking any cyber-physical attacker's unauthorized access with certainty. While such an assumption is quite common in the power system's literature, it may be regarded as questionable in the cyber-security context. In practice, the capabilities of cyber-physical attackers continuously evolve, cyber-security measures have specific protection scopes, while yet undiscovered (a.k.a., zero-day) vulnerabilities may always exist. In line with this, |hang \textit{et al.} \cite{zhang2022} consider the problem of allocating cyber-security devices that only reduce rather than completely eliminate the probability of a successful cyber-physical attack against the grid's transmission branches.
\color{black}

\subsection{Paper Scope \& contributions}

In this paper we also take the pragmatic viewpoint that cyber-security measures should not be considered as 100\% effective. On this basis, we investigate the stakes for an integrated \emph{cyber-physical risk management} framework wherein (i) preventive cyber-security measures are realistically considered to reduce rather than eliminate the probability of a successful cyber-physical attack, and (ii) the synergy with physical-security measures is acknowledged. To do so, we formulate a tri-level optimization problem seeking to choose in advance both physical- and cyber-security preventive measures, subject to uncertainty on the effectiveness of the latter. We model such uncertainty by way of a probability distribution of exogenous cyber-physical attackers differentiated in terms of their capability to overcome or not the cyber-security measures
protecting the power grid. Without loss in generality, we focus on the particular threat posed by cyber-physical attackers seeking to intrude into the power grid's substations and disconnect power generators and transmission branches (\textit{i.e.,} lines and transformers) so as to maximize operational costs, including the cost associated to involuntary load shedding. Anticipating such threat, a preventive planner may choose to update the security rules of the firewalls protecting the grid's substations and/or to procure reserve  that can be deployed in the aftermath of a successful attack. To solve the resulting problem, under standard assumptions, we rely on a modified version of the classic \emph{Column \& Constraint Generation} (C\&CG) algorithm by \cite{zeng2013solving}. Results obtained on two benchmark systems, namely the 24-bus version of the IEEE Reliability Test System \cite{grigg1999ieee} and the IEEE 118-bus system \cite{coffrin2019nestanictaenergytest}, showcase the complex interrelations between physical- and cyber-security measures.

\section{Assumptions, models \& methods}

\subsection{Problem description}

\begin{figure}[t]
    \centering
    \includegraphics[width=\linewidth]{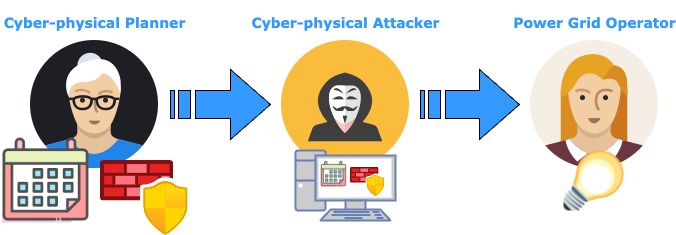}
    \caption{Cyber-physical power grid security actors}
    \label{fig:actors}
\end{figure}

We adopt a planning perspective and consider the problem of protecting the electric power system operation against the threat of malicious cyber-physical attacks.  Figure \ref{fig:actors} illustrates the independent decision-making actors involved in this problem. 

The central actor is the \emph{cyber-physical attacker} (a.k.a. \emph{attacker} for brevity) who  may seek to compromise the functionalities of the power grid's cyber sub-system so as to disrupt electricity supply. The left-most actor is the so-called \emph{cyber-physical planner} (a.k.a. \emph{planner} for brevity). Anticipating the threat posed by the cyber-physical attacker, she may act on (i) the power grid's cyber sub-system to ensure that it cannot be breached in a way causing an unacceptable electricity supply  and/or (ii) the physical power grid so as to ensure that it can maintain an acceptable electricity supply service in spite of the 
maloperation of its cyber sub-system. Finally, the right-most actor is the \emph{power grid operator} (a.k.a. \emph{operator} for brevity)  who, in the aftermath of any cyber-physical attack, must operate the (possibly compromised) electric power system.

We explicitly acknowledge that cyber-security measures have specific protection scopes and are also susceptible to unknown vulnerabilities (\textit{e.g.}, zero-day vulnerabilities), while the capabilities of   cyber-physical attackers are constantly evolving. In short, with any cyber-security measure comes the risk that a sufficiently capable cyber-physical attacker may still find a way to compromise the protected functionality. To account for this feature, we 
consider a discrete set of alternative cyber-physical attackers differentiated in terms of available resources (\textit{i.e.}, number of cyber sub-system components an attacker can compromise) and  capabilities to overcome the applicable cyber-security measures. We further associate a probability measure to this set so as to represent the threat landscape as a probability distribution over credible cyber-physical attackers.

Any credible cyber-physical attacker is assumed to be an independent decision-maker, choosing her attack vector upon observing the active cyber- and physical- security measures and while anticipating the optimal reaction of the power grid operator. The corresponding decision-making problem can be cast as a bi-level optimization problem. Similarly, the cyber-physical planner is considered to be an independent decision-maker choosing her risk management strategy while anticipating the optimal reaction of each alternative cyber-physical attacker.
Under information symmetry assumptions, the respective decision-making problem can be cast as a tri-level optimization problem under uncertainty. 

We consider a set-up wherein cyber-physical attackers may seek to intrude into digital substations\footnote{For ease of exposition, and without 
loss in generality, we henceforth assume that any single bus corresponds to a distinct digital substation and use the terms bus and substation interchangeably to describe the assumed threat.} and disconnect power generators and transmission lines  so as to maximize the power grid operational costs.  These latter costs include generation redispatching costs and costs of involuntary load shedding and are determined as per the optimal reaction of the power grid operator to the realized grid status. A cyber-physical attacker's strategy is constrained by a resource limit, specifying the number of substations she may simultaneously gain unauthorized access into. Once infiltrating a substation, the cyber-physical attacker can disconnect generators and transmission branches.

 Anticipating this uncertain threat, a cyber-physical planner may choose to update the rules of cyber-security devices blocking unauthorized access (a.k.a., firewalls) to the grid's substations. Additionally, this actor should also choose the system reserve capacity which constrains the redispatching of generating units in real-time operation. The objective of the   cyber-physical planner is to jointly minimize the cost of base-case operation (\textit{i.e.}, generation dispatch) and cyber-physical security measures (\textit{i.e.}, of updating firewall rules and reserve procurement) along with the expected cost of the power grid operation (\textit{i.e.}, expected generation redispatching and load shedding).

\subsection{Mathematical formulation}

Relying on the DC power flow approximation to keep the model of the grid physics linear, we express the trilevel optimization problem of the cyber-physical   planner as in (\ref{plan.obj.2} -- \ref{flow.bounds.2}). 

The top row of the planner's objective \eqref{plan.obj.2} includes all 1\textsuperscript{st}-stage costs, and specifically  the cost of the pre-attack generation dispatch, the cost of reserve procurement, and the cost of mobilizing resources  at the physical locations of the different buses of the system to update the firewall rules (3\textsuperscript{rd} term). These 1\textsuperscript{st}-stage costs are to be jointly minimized along with the expectation of the power grid operation cost over the post-attacks states while facing the threat described as the postulated set of cyber-physical attackers. The probability weighted summation of these costs is as shown in the bottom row of \eqref{plan.obj.2} and includes scenario-specific redispatching and load shedding costs. 

Inequality constraint \eqref{plan.resource.2} is a resource constraint limiting the total number of firewalls that the  cyber-physical planner can  update with new security rules in the considered  lead-time.  Constraints (\ref{plan.gen.bounds.2},\ref{gen.sign.2}) define the feasible region of the planner's generation variables, referring to the pre-attack dispatch and reserve capacity. Constraint \eqref{plan.bins.2} models the choice of where to update firewall rules as a binary decision. The group of constraints (\ref{plan.pbalance.2}
-- \ref{plan.flow.bounds.2}) enforces that the pre-attack dispatch implies no loss of load and conforms to the branch thermal ratings as per the DC power flow approximation. While the focus is on the considered cyber-physical threat, this group of constraints is included to ensure that reserve procurement is not unrealistic with respect to the pre-attack operation of the system.  

For any cyber-physical attacker $i \in \mathcal{A}\cup\mathcal{B}$ the objective function to be maximized is the cost of power grid operation, including attack-induced reserve deployment and load shedding \eqref{att.obj.2}. Constraint \eqref{attack.resource.2} defines an attacker's resources in terms of the number of buses she may simultaneously infiltrate.
Inequality constraints \eqref{att.def.2} are only valid for so-called cyber-physical attackers with \emph{basic capabilities} and enforce that any such actor may not infiltrate any bus that is protected by means of updated firewall rules\footnote{The complementary type of cyber-physical attacker that can overcome cyber-physical measures is referred to as an attacker with \emph{advanced capabilities}.}. The choice to intrude into the cyber sub-system of a specific bus is modeled through a binary decision variable \eqref{att.busbins.2}. Inequality constraints \eqref{att.branches.2} link the status of transmission branches and the potential intrusion of its associated (``sending/receiving'') end-buses. Specifically, these constraints enforce that a branch can only be disconnected if the cyber-physical attacker has managed to successfully intrude either its ``sending'' or ``receiving'' bus. We use binary decision variables to model branch disconnection decisions \eqref{att.branch.bins.2}. Equality constraints \eqref{att.gens.2}  model the disconnection of the generating units connected at a breached bus.

 Expressions (\ref{op.obj.2} -- \ref{flow.bounds.2}) model the operator's reaction to the realized attack and, under information symmetry assumptions, the impact of the attack on the power grid operation cost. Constraints (\ref{gen.up.bounds.2},\ref{gen.dn.bounds.2}) limit the reserve energy to be used by the   operator according to the capacity procured by the cyber-physical planner and, of course, the connection status of the respective unit. Equality constraints \eqref{pbalance.2} express the nodal power balance for the post-attack system state, featuring the additional decision variables corresponding to the deployment of reserve energy and the use of involuntary load shedding. Notice here the product of the pre-attack dispatch decided by the planner and the attacker's decision to compromise the concerned bus. In \eqref{load.shed.2}, the non-negative load shedding variable is upper bounded by the respective nodal load demand. Equality constraints \eqref{adv.flow.2} define the post-attack power flow as  the product of the branch disconnection variable and the fraction of the voltage angle difference divided by the branch reactance. Finally, inequalities \eqref{flow.bounds.2} impose the branch thermal ratings. 
\begin{align}
     \min_{\mathbf{p},\mathbf{r},\mathbf{z}}&\left\{\sum_{g \in \mathcal{G}}\left(C_{g}^{R} 
\cdot r_{g}+ C_{g}^D 
\cdot p_{g} \right)+ C^{Z} 
\cdot\sum_{  n \in \mathcal{N}}z_{n}\right. \nonumber\\
     &\left.+\sum_{i \in \mathcal{A}\cup\mathcal{B}}\pi_i 
\cdot \left(\sum_{g \in \mathcal{G}}C^{RD}_{g}
\cdot up_{g,i}+C^L 
\cdot \sum_{n\in \mathcal{N}}l_{n,i}\right)\right\},\label{plan.obj.2}\\
 &\sum_{n \in \mathcal{N}}z_{ n}\le Z,&\label{plan.resource.2}\\
 \intertext{\textit{for all generators} $g \in \mathcal{G}:$}
        &p_{g}+r_{g}\cdot \Delta T\le P_{g} \cdot \Delta T,&\label{plan.gen.bounds.2}\\
        &p_{g},r_{g}\ge 0, \label{gen.sign.2}\\ 
 \intertext{\textit{for all buses $n \in \mathcal{N}$}:}
    &z_{n} \in \{0,1\}  ,&\label{plan.bins.2}\\
  &\sum_{g \in \mathcal{G}} \gamma_{g,n}\cdot p_{g}=\sum_{t \in \mathcal{T}}\tau_{t,n}\cdot f_{t,0}+D_{n},& \label{plan.pbalance.2}\\
   \intertext{\textit{for all branches $t \in \mathcal{T}$}:}
    &f_{t,0}=\left(1/X_{t}\right) \sum_{n \in \mathcal{N}}\tau_{t,n} \theta_{n,0}, \forall t \in \mathcal{T},&\label{flow.def.2}\\
        & -F_{t} \le f_{t,0}\le F_{t}, &\label{plan.flow.bounds.2}\\
        \intertext{\textit{for all cyber-physical attackers} $i \in \mathcal{A}\cup\mathcal{B}$:}
 &\max_{\mathbf{v,w,y}}\left\{ \sum_{g \in \mathcal{G}}C^{RD}_{g}\cdot up_{g,i}+C^{L}\cdot \sum_{n\in \mathcal{N}}l_{n,i}\right\},&\label{att.obj.2}\\
&\hspace{8mm}  \sum_{n \in \mathcal{N}}(1-w_{n,i})\le W_{i},&\label{attack.resource.2}\\
   \intertext{\hspace{8mm}   \textit{for all buses $n \in \mathcal{N}$}:}
& \hspace{8mm}w_{n,i}\ge z_n,  \text{\textbf{iff}} \enspace i \in \mathcal{B},&\label{att.def.2}\\
&\hspace{8mm}  w_{n,i} \in \{0,1\},&\label{att.busbins.2}\\
   \intertext{\hspace{8mm}\textit{for all branches $t \in \mathcal{T}$}:}
&\hspace{8mm} v_{t,i}\ge \sum_{n \in \mathcal{N}}|\tau_{t,n}|\cdot w_{n,i} -1,&\label{att.branches.2}\\
&\hspace{8mm}v_{t,i} \in \{0,1\},&\label{att.branch.bins.2}\\
 \intertext{\hspace{8mm}   \textit{for all generators} $g \in \mathcal{G}:$}
 &\hspace{8mm}   y_{g,i}= \sum_{n \in \mathcal{N}}\gamma_{g,n}\cdot w_{n,i},&\label{att.gens.2}\\
  &\hspace{8mm}\min_{\mathbf{up},\mathbf{dp},\mathbf{l}}\left\{ \sum_{g \in \mathcal{G}}C^{RD}_{g}\cdot up_{g,i}+C^L\cdot \sum_{n\in \mathcal{N}}l_{n,i}\right\},&\label{op.obj.2}\\
   \intertext{\hspace{15mm} \textit{for all generators} $g \in \mathcal{G}:$}
       &\hspace{15mm} 0 \le up_{g,i}\le r_{g} \cdot \Delta T{\cdot y_{g,i}},&\label{gen.up.bounds.2}\\
        &\hspace{15mm} 0 \le dp_{g,i}\le p_{g}, &\label{gen.dn.bounds.2}\\
         \intertext{\hspace{15mm}\textit{for all nodes $n \in \mathcal{N}$}:}
  &\hspace{15mm}\sum_{g \in \mathcal{G}}\gamma_{g,n}\left(p_{g}{\cdot w_{n,i}}+up_{g,i}-dp_{g,i}\right)=&\nonumber\\
  &\hspace{25mm} \sum_{t \in \mathcal{T}}\tau_{t,n} f_{t,i}+(D_{n}-l_{n,i}),& \label{pbalance.2}\\
      &\hspace{15mm} 0 \le l_{n,i}\le D_{n},&\label{load.shed.2}\\
         \intertext{\hspace{15mm}\textit{for all branches $t \in \mathcal{T}$}:}
     &\hspace{15mm} f_{t,i}=\left(v_{t,i}/X_{t,i}\right)\sum_{n \in \mathcal{N}}\tau_{t,n} \theta_{n,i}, \label{adv.flow.2}&\\
             &\hspace{15mm} -F_{t} \le f_{t,i}\le F_{t},\forall t \in \mathcal{T},&\label{flow.bounds.2}\\
     \intertext{where $\Delta T=1 \enspace hr.$} \nonumber
\end{align}
\subsection{Algorithmic solution approach}
We rely on a modified version of the  \emph{Column \& Constraint Generation} (C\&CG) by \cite{zeng2013solving} to solve tri-level optimization problem (\ref{plan.obj.2} -- \ref{flow.bounds.2}). To the best of our knowledge,  \cite{yuan2014optimal} first applied the C\&CG algorithm in the context of 
 electric power systems security management to solve the \emph{deterministic} optimization problem of choosing which  transmission branches to protect facing  an attacker with the ability to physically disconnect transmission branches. \cite{xiang2018improved} used a modified version of this algorithm in a stochastic setting,  considering  the resources of the attacker as uncertain.
 
In order to briefly explain the C\&CG algorithm, let us first notice that constraints (\ref{attack.resource.2},\ref{att.busbins.2} -- \ref{att.gens.2}) \emph{implicitly} form a finite \emph{attack surface} consisting of distinct attack vector scenarios (\textit{i.e.}, substations to intrude and associated branches \& generators to disconnect) per cyber-physical attacker $i \in \mathcal{A}\cup\mathcal{B}$.  We will henceforth denote the attack surface for any single cyber-physical attacker as $\mathcal{S}_i$. In problem (\ref{att.obj.2} --\ref{flow.bounds.2}), any cyber-physical attacker essentially chooses the attack vector  with the maximal impact on the power grid operation cost out of her corresponding finite set $\mathcal{S}_i$. This choice of course depends on the cyber-physical planner's decisions protecting the power grid operation.  These latter decisions in turn depend on the attack vectors that the cyber-physical planner anticipates the alternative attackers to potentially launch. The idea then is to iteratively
update the planner's decisions 
 while generating for each alternative cyber-physical attacker $i$ a growing sequence ${\mathcal{S}}_i^k$ of subsets containing  maximal-impact scenarios chosen from the attack vectors $\mathcal{S}_i$ that can be launched by this attacker. In other words, \emph{explicitly} form finite subsets of attack vector scenarios that would allow the planner to make ``similar'' decisions as the decisions that would be made while facing the complete attack surface $\cup_{i \in \mathcal{A}\cup\mathcal{B}}\mathcal{S}_i$.  By way of ``similar'' decisions, we refer here to near-optimal decisions that imply practically equal total costs (\textit{i.e.,} base case dispatch costs, reserve procurement costs and expected impact of the cyber-physical attack on the power grid operation). 

\begin{figure}
    \centering
    \includegraphics[width=\linewidth]{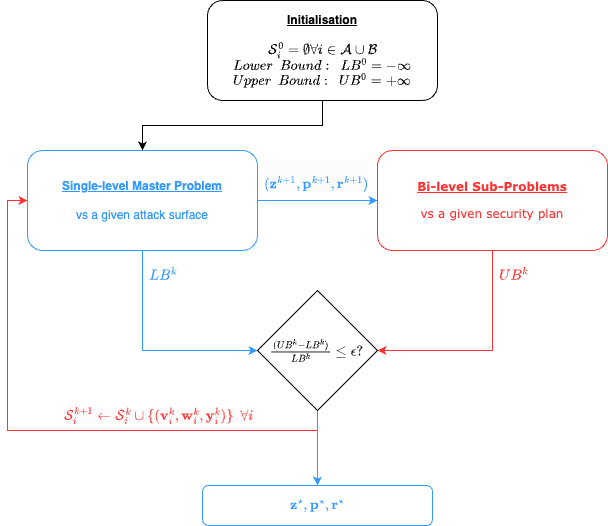}
    \caption{Column \& constraint generation flowchart}
    \label{fig:ccg}
\end{figure}

Figure \ref{fig:ccg} presents the flowchart of the implemented CC\&G algorithm. In every iteration $k$,  given the current finite subsets $\mathcal{S}_{i}^k$, a  \emph{Single-level Master Problem} is solved to update the decisions of the cyber-physical planner regarding the updating of firewall rules ($\mathbf{z}^k$) as well as the dispatch and reserve capacity of generating units ($\mathbf{p}^k,\mathbf{r}^k$). The solution of this  problem also provides an updated lower bound for the total cost which is optimized while taking into consideration subsets $\mathcal{S}_{i}^k$. Given the updated decisions of the cyber-physical planner ($\mathbf{z}^k,\mathbf{p}^k,\mathbf{r}^k$), we then solve a \emph{Bi-level Sub-Problem} per cyber-physical attacker $i \in \mathcal{A}
\cup\mathcal{B}$ to update her attack vector ($\mathbf{v}^k_i,\mathbf{w}^k_i,\mathbf{y}^k_i$). Along with the updated attack vectors, the resolution of all bi-level problems  also allows to compute an updated estimation for the 2\textsuperscript{nd} stage costs expressing the expected impact of the potential attacks. We add this updated estimation to the cost of the updated 1\textsuperscript{st} stage decisions ($\mathbf{z}^k,\mathbf{p}^k,\mathbf{r}^k$) to update an upper bound for the total cost. While the difference between the upper and lower bounds is not within the specified tolerance, the attack surface of any alternative cyber-physical attacker is updated as in $\mathcal{S}_{i}^{k+1}\leftarrow\mathcal{S}_{i}^k\cup \{(\mathbf{v}_i^k,\mathbf{w}_i^k,\mathbf{y}_i^k)\}$.

\subsubsection{Single-level Master Problem: \textit{vs} a given attack surface}

At every iteration of the algorithm, and given an updated attack surface $\cup_{i \in \mathcal{A}\cup\mathcal{B}}\mathcal{S}_i^k$, we formulate a single-level scenario sub-problem that corresponds to tri-level problem (\ref{plan.obj.2} -- \ref{flow.bounds.2}). To do so, we first introduce attack-vector-specific 2\textsuperscript{nd} stage variables ${up}^{s}_{g,i}$ and ${l}^s_{n,i}$ per cyber-physical attacker $i$ and attack vector $s$ and then replace the 2\textsuperscript{nd} row of \eqref{plan.obj.2}
with: 
\begin{align}
\left. + \sum_{i \in \mathcal{A}\cup\mathcal{B}}\pi_i \cdot \eta_i\right\},&\nonumber
\end{align}
and the maximization operation \eqref{att.obj.2} with:
\begin{align}\eta_i &\ge     \sum_{g \in \mathcal{G}}C^{RD}_{g}\cdot {up}^{s}_{g,i}+C^{L}\cdot \sum_{n\in \mathcal{N}}{l}^s_{n,i}  \enspace \forall i \in \{\mathcal{A}\cup\mathcal{B}\}, \forall s \in \mathcal{S}_i^{k}. \nonumber
\end{align}

Constraints (\ref{attack.resource.2} -- \ref{att.busbins.2}, \ref{att.gens.2}, \ref{gen.up.bounds.2} --\ref{load.shed.2},\ref{flow.bounds.2}) can be directly re-written per cyber-physical attacker and attack vector. Further, the operator's objective function \eqref{op.obj.2} can be omitted since this actor and the cyber-physical planner are cooperating towards the same interest. The only noteworthy aspect is the treatment of transmission branch constraints (\ref{att.branches.2},\ref{att.branch.bins.2}) along with \eqref{adv.flow.2}. We recall here that (\ref{att.branches.2},\ref{att.branch.bins.2}) allow a cyber-physical attacker to disconnect  a transmission branch provided that she has intruded into a substation including any one of its end-buses. In the single-level reformulation, and while considering the finite attack vector scenario subsets $\cup_{i \in \mathcal{A}\cup\mathcal{B}}\mathcal{S}_i^k$, we must enforce that any transmission branch under attack will be disconnected if any one of its end-buses is attacked and not successfully protected (either by relying on outdated firewall rules or by facing an attacker that can by-pass the updated rules). To do so, we use the following expressions:
\begin{align}
        \intertext{\textit{For all cyber-physical attackers with \textbf{basic} capabilities} $i \in \mathcal{B}$, \textit{attack vector scenarios} $s \in \mathcal{S}_i^k$ \textit{and \textbf{attacked} transmission branches} $t \in \mathcal{T}: v_{t,i}^s=0$:}
    f_{t,i}^s &\le M \cdot w_{n_{s}(t),i}^s \label{ref.flow.s1}\\
    f_{t,i}^s &\ge -M \cdot w_{n_{s}(t),i}^s \label{ref.flow.s2}\\
       f_{t,i}^s &\le M \cdot w_{n_{r}(t),i}^s \label{ref.flow.r1}\\
    f_{t,i}^s &\ge -M \cdot w_{n_{r}(t),i}^s\label{ref.flow.r2}\\
     f_{t,i}^s&\le \left(v_{t,i}^s/X_{t,i}\right)\sum_{n \in \mathcal{N}}\tau_{t,n} \theta_{n,i}^s +M\cdot (2-w_{n_{s}(t),i}^s +w_{n_{r}(t),i}^s), \label{ref.floweq.1}&\\
      f_{t,i}^s&\ge \left(v_{t,i}^s/X_{t,i}\right)\sum_{n \in \mathcal{N}}\tau_{t,n} \theta_{n,i}^s +M\cdot (w_{n_{s}(t),i}^s +w_{n_{r}(t),i}^s-2) \label{ref.floweq.2},&\\
       \intertext{\textit{for all cyber-physical attackers with \textbf{advanced} capabilities} $i \in \mathcal{A}$, \textit{attack vector scenarios} $s \in \mathcal{S}_i^k$ \textit{and \textbf{attacked} transmission branches} $t \in \mathcal{T}: v_{t,i}^s=0$:}
       f_{t,i}^s&=0,\label{ref.flow.adv}\\
       \intertext{\textit{for all cyber-physical attackers} $i \in \mathcal{A}\cup\mathcal{B}$, \textit{attack vector scenarios} $s \in \mathcal{S}_i^k$ \textit{and \textbf{non-attacked} transmission branches} $t \in \mathcal{T}: v_{t,i}^s=1$:}
        f_{t,i}^s&=\left(1/X_{t,i}\right)\sum_{n \in \mathcal{N}}\tau_{t,n} \theta_{n,i}^s. \label{ref.naflow}&
\end{align}

In (\ref{ref.flow.s1} -- \ref{ref.floweq.2}) indices $n_{s}(t),n_{r}(t)$
denote the sending, receiving bus of transmission branch $t$. Accordingly, inequality constraints (\ref{ref.flow.s1} -- \ref{ref.flow.r2}) set the power flow through a branch to zero if at least one of its end-buses is successfully attacked.  Inequalities (\ref{ref.floweq.1} -- \ref{ref.floweq.2}) enforce the power flow definition from equality \eqref{adv.flow.2} only for branches whose end-buses are both protected successfully. Equality constraint \eqref{ref.flow.adv} only holds for the sub-class of cyber-physical attackers with advanced capabilities. Since these actors can by-pass the cyber-security measures, the power flow of any attacked branch will in any case be equal to zero. Finally, the standard DC power flow definition equality  is imposed for any transmission branch that is not under attack \eqref{ref.naflow}.

\subsubsection{Bi-level Sub-Problems: \textit{vs} a given cyber-physical security plan}
At every iteration of the algorithm, and given  updated decisions for the cyber-physical planner ($\mathbf{z}^k,\mathbf{p}^k,\mathbf{r}^k$), we formulate per cyber-physical attacker $i\in \mathcal{A}\cup\mathcal{B}$ a bi-level problem as in (\ref{att.obj.2} -- \ref{flow.bounds.2}). To solve these problems we rely on the fact that  the lowest-level minimization problem (\ref{op.obj.2} -- \ref{flow.bounds.2}) is a convex optimization problem. Exploiting this property, we remove the minimization operator of \eqref{op.obj.2}, introduce the dual feasibility constraints of (\ref{op.obj.2} -- \ref{flow.bounds.2}) and enforce the strong duality condition \cite{duality}.

\section{Results}
\label{results}

\begin{figure}[]
    \centering
    \includegraphics[width=0.6\linewidth]{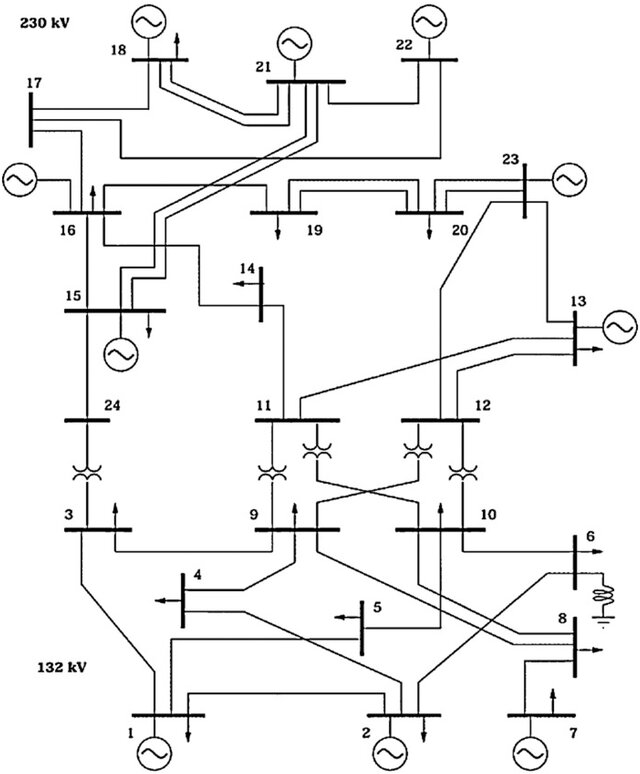}
    \caption{Single-line diagram of the IEEE 24-bus benchmark \cite{24-diagram}}
    \label{fig:24-bus}
\end{figure}

\begin{figure}[h]
    \centering
    \includegraphics[width=\linewidth]{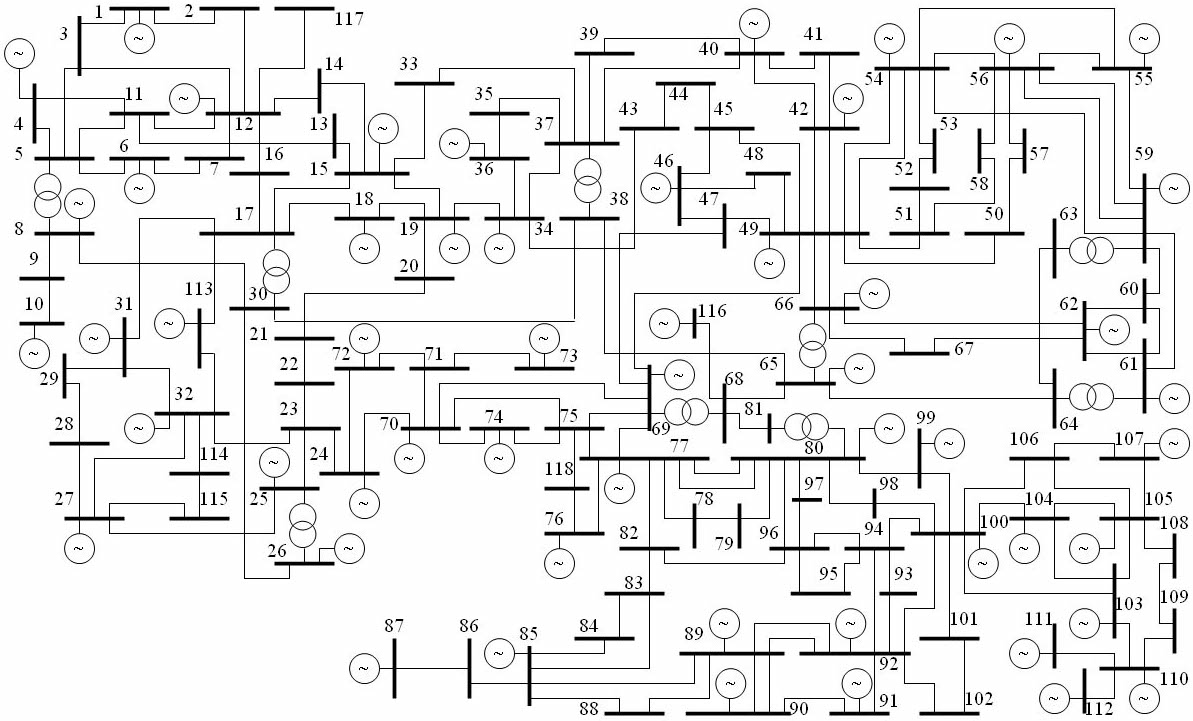}
    \caption{Single-line diagram of the IEEE 118-bus benchmark \cite{Al-Roomi2015}}
    \label{fig:118-bus}
\end{figure}

We consider a set of case studies defined on two well known benchmarks, namely the IEEE 24-bus and 118-bus systems. Figures \ref{fig:24-bus} and \ref{fig:118-bus} plot the respective single-line diagrams. Our motivation for using two different benchmarks is to derive observations that do not depend on the properties of the considered system. We use all original system data as documented by \cite{coffrin2019nestanictaenergytest} and supplement these with:
\begin{enumerate}
    \item An assumed \emph{value of lost load} parameter, set at 5000 monetary units per MWh;
    \smallskip
    \item an assumed \emph{cost of updating firewall rules} parameter, set at 5.55 monetary units per device;
    \smallskip
    \item An assumed \emph{reserve capacity cost} parameter, set per generator at 25\% of the respective dispatch cost parameter.
\end{enumerate}

To put results into context, we will henceforth report all cost data in percentage of the base-case dispatch cost of the respective benchmark. Such cost is computed while considering an empty set of potential cyber-physical attackers and solving optimization problem (\ref{plan.obj.2} -- \ref{plan.flow.bounds.2}). It amounts to 41904.11 monetary units for the IEEE 24-bus benchmark and 109826.08 monetary units for the IEEE 118-bus benchmark. Table \ref{mycases} collects the test cases considered and introduces the order in which results will be presented. As can be seen, apart from different benchmark systems, we differentiate test cases in terms of the cyber-security resources available to the planner (3\textsuperscript{rd} column), the resources available to a cyber-physical attacker with basic (\textit{BSC}) and/or advanced (\textit{ADV}) capabilities (4\textsuperscript{th} and 5\textsuperscript{th} columns respectively) as well as the probabilities associated to these actors (6\textsuperscript{th} and 7\textsuperscript{th} columns respectively).


\begin{table}
      \centering
    \begin{tabular}{c c| cc   c| c c}
    \sphline
    \multirow{3}{*}{Case} & 
    \multirow{3}{*}{System} & \multicolumn{3}{|c|}{Resources} &\multirow{2}{*}{\hspace{3mm} Attack}&\multirow{2}{*}{\hspace{-3mm}Probability} \\
    &&Planner& \multicolumn{2}{c|}{Attackers}\\
    &&&\textit{BSC}& \textit{ADV}&\textit{BSC}& \textit{ADV}\\
    \sphline
    A & 24-bus&24&2&0&0.01&0\\
    B & 24-bus&24&0&2&0.0 & 0.01\\
    C & 24-bus&24&2&2&0.005 & 0.005\\
    \sphline
     D & 24-bus&3&2&0&0.01&0\\
    E & 24-bus&3&0&2&0.0 & 0.01\\
    F & 24-bus&3&2&2&0.005 & 0.005\\
    \sphline
         G & 118-bus&3&2&0&0.01&0\\
    H & 118-bus&3&0&2&0.0 & 0.01\\
    I & 118-bus&3&2&2&0.005 & 0.005\\
    \sphline
    \end{tabular}
    \caption{Summary of test case parameters}
    \label{mycases}
\end{table}

\subsection{Unlimited cyber-security resources}

We start by considering a set of test-cases on the 24-bus system and while assuming that the cyber-physical planner has unlimited cyber-security resources that would allow her to update the firewalls rules for all buses of the system. We first consider in case A that there exists only a single credible cyber-physical attacker and this actor does not have the capability to overcome the applicable cyber-security measures. In complementary case B we consider that the same cyber-physical attacker can by-pass firewalls, even under the updated rules. In both cases, the probability associated to the single credible attacker is set at 0.01 and her resource limit is set to 2, meaning that she has the ability to intrude at most 2 buses. Table \ref{tab:cases_ABC} includes the results for these two test cases.
 \begin{table}[h]
      \centering
    \begin{tabular}{c| c | c c c}
        \sphline
        \multirow{2}{*} {Case}&Buses& \multicolumn{3}{|c}{Costs (\%)} \\
        &Secured&Total & 1\textsuperscript{st}-stage  & Exp.(2\textsuperscript{nd}-stage)  \\ 
         \sphline
         A & 1 -- 10, 12 -- 23& 100.29& 100&0\\
         B &  --  & 200.54& 117.17&83.37\\
        C & 1 -- 10, 13 -- 23& 157.35	&	107.20 &	49.88\\
         \sphline
    \end{tabular}
    \caption{Solution overview for test cases A -- C}
    \label{tab:cases_ABC}
\end{table}

Starting from the data in the top row of Table \ref{tab:cases_ABC}, if cyber-security measures are fully effective, in the assumed conditions it makes (economic) sense to fully rely on these so as to completely neutralize the threat posed by the cyber-physical attacker. Indeed, the optimal strategy here is to pay a negligible additional cost of 0.29\% at the planning stage so as to ensure that a potential cyber-physical attack may not have any impact on the power grid operation. Notice that this additional 0.29\% is only spent on updating the firewall rules, since the optimal costs of generation dispatch and reserve procurement are  identical to the aforementioned base-case results (computed while assuming no credible threat). It should also be noted that the optimal planning strategy is to deploy the considered cyber-security measure in 22 out of the 24 system buses. Recalling here that the 24-bus benchmark is relatively lightly loaded and has ample transmission capacity, in the considered steady-state, linear model it could withstand the outage of two transmission (\textit{i.e.}, no load or generation) buses.

The middle row of Table \ref{tab:cases_ABC} concerns the case wherein the cyber-physical attacker would have the advanced capability to by-pass the cyber-security measures under consideration. It is easy to understand that, facing only such a threat actor, there is no point in using cyber-security measures. Rather, the only form of security can come from physical measures. In the assumed conditions, it makes (economic) sense to spend an additional cost of 17.17\% in choosing the generation dispatch and procuring reserve capacity. Still, this would not completely neutralize the credible threat which carries a substantial expected impact. It would be equal to 83.37\% of the base-case dispatch cost. Our detailed results indicate that the cyber-physical attacker's strategy would be to intrude buses 15 (load and significant generation capacity) and 23 (largest generation capacity) thus provoking both reserve deployment and involuntary load shedding.  

Let us now move on to test case C, which as seen in Table \ref{mycases} concerns two alternative credible cyber-physical attackers differentiated only in terms of their ability to overcome or not the updated firewall rules. Both these actors have the same resources as in cases A and B and share equally the total probability associated to the attackers in cases A or B. The bottom row of Table \ref{tab:cases_ABC} reports the corresponding solution overview. As evidenced by these results, acknowledging that a cyber-physical attacker  may not be impeded by cyber-security measures 
results in using a blend of cyber- and physical- security measures. Indeed, the optimal solution includes both updating the firewall rules for 21 system buses and paying an additional 7.2\% in generation dispatch and reserve capacity. What is most interesting here is the synergy between the latter and former measures. The physical system is more strongly prepared, by way of the generation dispatch and reserve capacity, since possibly facing an attacker that could by-pass cyber-security measures. This strength also carries over to possibly facing the alternative attacker that would not be able to by-pass the firewall rules.  The available reserve capacity, along with the reduced probability of the cyber-physical attacker with basic capabilities, imply that she is expected to only cause a minor impact on the power grid operation. As a result, the security of the cyber sub-system  against this threat can be slightly relaxed. Compared to case A, one less bus is chosen to be protected by cyber-security measures in case C.

\subsection{Limited cyber-security resources}

We continue with a set of test-cases on the 24-bus and 118-bus benchmarks in the setting where the cyber-security resources of the  planner are limited. We believe that considering such limitation is very relevant, since in practice deploying cyber-security measures  may involve mobilizing specialized workforce  to access the concerned systems locally. For both benchmarks, we limit the cyber-physical planner's resources to 3 while maintaining the assumption that any attacker could at most intrude into 2 buses. We form alternative test cases by varying the sets of credible cyber-physical attackers in the same style as previously. Specifically, for any benchmark, we consider the alternative hypotheses of  facing (i,ii) a single credible attacker with either basic or advanced capabilities as well as (iii) two credible attackers with basic and advanced capabilities respectively.
\subsubsection{24-bus benchmark}
\begin{table}[h]
      \centering
    \begin{tabular}{c| c | c c c}
        \sphline
        \multirow{2}{*} {Case}&Buses  & \multicolumn{3}{|c}{Costs (\%)} \\
        &Secured&Total &  1\textsuperscript{st}-stage  & Exp.(2\textsuperscript{nd}-stage)  \\ 
         \sphline
         D & 15,18,23& 172.47& 117.54&54.89\\
         E & -- & 200.54& 117.17&83.37\\
         F &  15,18,23  & 186.91& 117.73&69.13\\
         \sphline
 \end{tabular}
    \caption{Solution overview for test cases D -- F}
    \label{tab:cases_DEF}
    \end{table}

Table \ref{tab:cases_DEF} concerns the test cases on the 24-bus benchmark. The results for case D, compared to those for case A, confirm that the limit on the available cyber-security resources is indeed restrictive. This restriction is partly counterbalanced by the procurement of reserve capacity at an additional cost of 17.54\%. Once again we can notice here how cyber- and physical- security measures can complement one another if chosen in an integrated manner. It would however be impossible to completely neutralize the impact of the considered cyber-physical attack on the power grid operation cost, since the deployment of the procured reserve also comes at a cost. It is also interesting to acknowledge that the cyber-physical planner's optimal decision making includes updating the firewall rules for buses 15 and 23. We recall from case B that these would form the optimal attack vector for the attacker with advanced capabilities. The results for case E are, as anticipated, identical to those of case B. Since both these cases concern an attacker with advanced capabilities only, the cyber-security measures are ineffective hence the respective resource limit is immaterial. Comparing results for cases F and D, we find that, for this specific benchmark, the consideration that a cyber-physical attacker may find a way to overcome the applied cyber-security measures has a moderate effect on the optimal decisions of the planner regarding physical security measures and no effect on the optimal decisions regarding cyber-security measures. Indeed, the  only difference is in the optimal reserve \& dispatch cost which is only increased by 0.19\%. It appears preferable, to tolerate a larger expected impact of a potential attack on the power grid operation cost, and precisely increased by  14.24\%.

\subsubsection{118-bus benchmark}

\begin{table}[h]
      \centering
    \begin{tabular}{c| c | c c c}
        \sphline
        \multirow{2}{*} {Case}&Buses  & \multicolumn{3}{|c}{Costs (\%)} \\
        &Secured&Total & 1\textsuperscript{st}-stage  & Exp.(2\textsuperscript{nd}-stage)  \\ 
         \sphline
         G & 26,59,80& 129.05& 108.20&20.83\\
         H &  --  & 136.78& 106&30.78\\
         I &  26,59,69  & 133.99& 107.52&26.45\\
         \sphline
    \end{tabular}
    \caption{Solution overview for test cases G -- I}
    \label{tab:cases_GHI}
    \end{table}
    
Table \ref{tab:cases_GHI} concerns the test cases on the 118-bus benchmark. We make an interesting observation by comparing the outcomes for cases G and I. Recall that case G  only acknowledges  the possibility of a cyber-physical attacker with basic   capabilities while case I also acknowledges the possibility of an attacker with the advanced capabilities  to overcome  cyber-security measures. As per our results, the optimal choice of which buses to protect by means of updated firewall rules is different between these two cases. Again, the reason is the synergy between cyber- and physical- security measures. The hypothesis in case I calls for different physical-security measures, serving as  the optimal way to counterbalance the threat posed by the cyber-physical attacker with advanced capabilities. These physical measures would also be effective for the alternative of the attacker with basic capabilities, providing a form of security and leading to a different choice of (complementary) cyber-security measures.
 
\begin{table}[h]
      \centering
    \begin{tabular}{c|c |c |c|  c| c }
        \sphline 
   \multirow{3}{*}{Iteration}  &   \multicolumn{3}{c|}{Buses} &  \multicolumn{2}{c}{Cost (\&)} \\
              & \multirow{2}{*}{Secured} & \multicolumn{2}{|c|}{Attacked}  & Reserve &\multirow{2}{*}{Total}\\
              &     & Basic &Advanced & \& Dispatch& \\
         \sphline
1&&69,80&69,80& 100&100\\
2&69,80&10,26&66,80&108.39&116.87\\
3&10,26,80&69,66&69,66&108.69&117.32\\
4&66,10,80&59,49&59,49&108.76&119.36\\
5&69,59,10&66,100&69,100&106&124.46\\
6&69,59,10&26,9&69,80&108.61&129.24\\
7&69,59,26&89,80&69,80&108.61&129.24\\
8&59,26,80&69,89&69,80&109.39&129.41\\
9&59,26,80&100,89&59,49&109.33&132.34\\
10&26,49,80&66,59&59,49&109.82&133.35\\
11&59,26,80&49,8&59,49&109.82&133.35\\
12&69,59,26&68,66&69,80&107.16&133.57\\
13&59,26,89&100,80&59,49&108.04&133.68\\
\sphline
14&69,59,26&10,80&69,80&107.52&133.99\\
\sphline
    \end{tabular}
    \caption{CC\&G algorithm evolution for test case I}
    \label{tab:algo_i}
\end{table} 
Table \ref{tab:algo_i} details the evolution of the problem optimal solution as per the C\&CG algorithm for test case I. Per iteration, the 2\textsuperscript{nd} column of this table holds the indices of the buses wherein the cyber-physical planner would decide to update the firewall rules. The 3\textsuperscript{rd} and 4\textsuperscript{th} columns hold the attack vectors launched by the cyber-physical attackers with basic and advanced capabilities in response to the planner's updated decisions. The last two columns show the cost of the optimal physical security measures and the total cost of the cyber-physical planner (\textit{i.e.}, her objective).

At the 1\textsuperscript{st} iteration of the algorithm, the cyber-physical planner considers no credible threat and thus implements no security measure. Both attackers would in such case launch the same attack vector since none of them is impeded by any barrier. Acknowledging this possibility, at the 2\textsuperscript{nd} iteration, the cyber-physical planner chooses to both (i) completely neutralize the attack vector of the attacker with basic capabilities (by updating firewall rules) and (ii)   mitigate the potential impact of the alternative attacker with advanced capabilities (by changing the generation dispatch and procuring reserve capacity). It is also interesting to comment on the solution for iteration 4. Since the planner's resource limit is set at 3, she may not completely neutralize the subset of attack vectors for the attacker with basic capabilities. This subset already includes 3 alternative pairs of buses and therefore requires 5 resources to be rendered ineffectual.
Notice that bus 69 is common in two of these attack vectors. The planner's optimal decision is to spread her resources to partially mitigate each one of these 3 alternative vectors rather than completely neutralize one of them. Doing so, the planner is seeking to ensure that there is no attack vector with considerably high impact that would also push the expectation term and the overall objective higher. It is also interesting that bus 69 is  not one of the buses to be protected by the available cyber-security resources. We found that this bus would also be targeted by the cyber-physical attacker with advanced capabilities and infer that physical security measures have been chosen to mitigate this threat.

\color{black}

It is also obvious from Table \ref{tab:algo_i} that, as anticipated, the cyber-physical attacker with advanced capabilities only reacts to an update of the physical security measures. Indeed, between iterations where these measures are not updated this actor would repeatedly launch the same vector (\textit{e.g.,} between iterations 6 and 7). We should explain here that it is not necessary for the planner to update both cyber- and physical- security measures at every iteration, since seeking to co-optimize her planning cost along with the \emph{joint} expectation of the attack impact over the set of credible attackers. We also notice that the cyber-physical attacker with advanced capabilities would frequently return to the same attack vector (buses 69,80) even when the physical- security measures seem to change. The updated dispatching and reserve decisions reduce the expected impact of this vector, however it still remains the best choice for this attacker relatively to her other options. We finally remark that while the total cost for the cyber-physical planner can only increase between different iterations, the optimal mix between 1\textsuperscript{st} and 2\textsuperscript{nd} stage costs is not necessarily constant. This explains why the reserve \& dispatch cost appears to decrease between some successive iterations.

\color{black}

\subsection{Computational Cost}
Our implementation of the C\&CG algorithm for the resolution of tri-level optimization problem  (\ref{plan.obj.2} -- \ref{flow.bounds.2}) was written in Julia \cite{julia}, using the JuMP modeling language \cite{jump}. We used the  PowerModels.jl  package \cite{PowerModels} for parsing the benchmark data and solved all optimization problem instances with the CPLEX solver \cite{cplex}. All test cases where solved using a MacBook with an 8-core M2 chip and 24 GB of RAM. 
\begin{table}[h]
      \centering
    \begin{tabular}{c|c cc c| c}
        \sphline
       \multirow{3}{*} {Case}& \multicolumn{4}{c|}{Execution time (sec)} & 
        \multirow{3}{*}{Iterations}\\
        & \multirow{2}{*}{Total}& \multirow{2}{*}{Master} &  \multicolumn{2}{c|} {Sub-problems}& \\
        &      & & Basic & Advanced &\\
        \sphline
        A & 8.38& 1.1&7.28&0&14\\
        B& 2.06&0.61&0&1.45&3\\
        C& 13.52&1.24&7.40&4.87&13\\
        D&8.41&1.62&6.79&0&16\\
        E&2.06&0.61&0&1.45&3\\
        F&10.57&1.14&4.61&4.82&11\\
        G&1409.24&9.24&1400&0&15\\
        H&605.4&7.56&0&597.84&6\\
        I&3307.55&20.75&1600.80&1686&14\\
        \sphline
      \end{tabular}
         \caption{Computational Cost}
    \label{tab:comp_cost}
    \end{table}

Table \ref{tab:comp_cost} reports the computational cost of all test cases in terms of the total execution time and number of iterations. The effect of the size of the considered system is obvious. We may also notice by comparing the data for cases A -- C to those 
for cases D -- F that the assumed resource constraint of the cyber-physical planner has a marginal effect on computational cost. On the other hand, we have found that 
the assumed value for a cyber-physical attacker's resource constraint has a multifold effect on the computational cost. Firstly, it generally increases the number of iterations required to achieve convergence. As the algorithm requires more iterations to converge to the optimal solution, the \emph{Master Problem} also requires more execution time per iteration to be solved. Finally, marginally increasing the resource 
limit also increases the execution time required to solve  each instance of the \emph{Bi-level Sub-problem}  (\ref{att.obj.2} -- \ref{flow.bounds.2}). Table \ref{tab:comp_cost_sens}  demonstrates these effects by documenting the computational cost of variants of case C with additional resources available to both alternative cyber-physical attackers.

\begin{table}[h]
      \centering
           \begin{tabular}{c| cc c| c}
        \sphline
       \multirow{3}{*} {Attacker Resources}& \multicolumn{3}{c|}{Execution time/iteration (sec)} & 
        \multirow{3}{*}{Iterations}\\
        & \multirow{2}{*}{Master} &  \multicolumn{2}{c|} {Sub-problems}& \\
                 & & Basic & Advanced &\\
        \sphline
        2& 0.10&0.42&0.44&11\\
        4&0.17&0.99&1.37&15\\
        6&1.01&1.78&1.65&30\\
        \sphline
        \end{tabular}
      \caption{Sensitivity of the computational cost to the attacker resources}
    \label{tab:comp_cost_sens}
    \end{table}
    
\color{black}
\section{Conclusions}

In this paper we focused on the threat posed by (malicious) cyber-physical attackers, who may use the cyber sub-system of the power grid as the medium to disrupt the continuous supply of electricity. The topic is rather timely not only due to the increasing role of digital monitoring, control, communications and computations infrastructures in the operation of the modern power grid but, unfortunately, also due to geo-political developments. The salient feature of this work is the acknowledgment that even though cyber-security measures can be  effective, a sophisticated threat actor may  eventually find a (new) way to circumvent these. We therefore investigated the stakes for an integrated risk management approach choosing both cyber- and physical- security measures under uncertainty on the effectiveness of the former.

To do so, we formalized an optimization problem from the perspective of a cyber-physical planner seeking to co-optimize her costs along with the expected impact of a cyber-physical attack on the power grid operation cost. The latter includes a cost of energy not supplied component, penalizing the event of an attack leading to electricity supply interruption. The planner faces a distribution of alternative credible cyber-physical attackers, including  attackers that can overcome cyber-security measures. Our main finding is a dual synergy between cyber- and physical security measures.  The use of physical- security measures was shown to be  primarily associated to the potential ineffectiveness of cyber-security measures. Put simply, an attacker who may overcome cyber-security measures would be counteracted by preparing the physical grid to withstand her attack vector. As a secondary effect, taking physical security measures against this  credible threat 
makes the grid physically stronger,
thus more able to withstand any credible threat. In fine, this physically stronger grid may have different vulnerabilities and different priorities/needs for cyber-security. There is therefore scope for  choosing cyber- and physical- security measures in an integrated manner.

It is relevant to mention here the distinctive features of the considered threat with respect to the `classical', \emph{physical-only} threats to the security of the electric power grid, such as component outages (a.k.a. contingencies) and power injection forecasting errors. While physical-only threats are purely exogenous and show relatively favorable statistical properties, cyber-physical threats are adaptive to the security management strategy and intentionally designed for the maximum attainable impact. To represent these features mathematically, we adopted the framework of tri-level optimization (under uncertainty).  Admittedly, the solvability of our tri-level optimization framework comes with restrictions that we have imposed on the representation of the grid physics and of its real-time control problem statement (steady-state and linear power flow model, with only continuous real-time controls and no infeasibility). These restrictions however do not seem critical for the conceptual question we address here and the qualitative conclusion we reach.  It is nevertheless desirable to overcome these restrictions in order to develop more realistic solution strategies to more advanced versions of the integrated cyber-physical power grid security management problem.

We remark here that these solvability restrictions mostly refer to the lowest-level optimization problem  (\ref{op.obj.2} -- \ref{flow.bounds.2}) used by a cyber-physical attacker so as to anticipate the power grid operator's optimal reaction. One may argue that in practice cyber-physical attackers are equally bound by similar solvability restrictions and would have to resort to similar simplifying assumptions in order to identify and launch near optimal attack vectors. In this spirit, a possible next research step would be to differentiate between the physical model used in (\ref{op.obj.2} -- \ref{flow.bounds.2}) to determine  the optimal attack vector(s) of the considered cyber-physical attacker(s) and the physical model used to assess and control the impact of such attack vector(s) on the power grid. Indeed, there is no \emph{a priori} restriction in using a more detailed physical model neither in the so-called \emph{Master Problem} nor in an additional post-processing step. This additional step would serve to evaluate more precisely an \emph{upper bound} given the resolution of all \emph{Bilevel Sub-problem(s)} and the respective attack vector(s).
 
The computational cost of the tri-level optimization framework is clearly also an issue that requires dedicated attention and further research. In spite of such cost, we relied on this framework  since it offers the transparency/interpretability required for the analysis we sought to perform here. Several improvements in algorithmic efficiency are conceivable on top of the basic implementation that we used, including for instance the resolution of Bi-level Sub-problem(s) in parallel over the credible set of cyber-physical attackers. We wish to also acknowledge the potential interest in moving away from  approaches purely based on analytical optimization techniques. The particular alternative approach that we aim to explore in future research is \emph{Multi-agent Reinforcement Learning}. We refer the interested reader to \cite{li2019reinforcement} for a comprehensive exposition of the relevance of using  Reinforcement Learning for cyber-physical system security management.

As a concluding point, we return to the original question and the integration of cyber- and physical- resources for electric power system security management. We have shown here that the integration of such resources is indeed well-justifiable, especially given the risk that malicious adversaries may develop the skill to overcome cyber-security measures. We showed this while focusing on a single instance of power  power grid operation setting and on cyber- and physical- measures that could conceivably be chosen on similar time-frames. Addressing the electric-power system cyber-physical security management problem as a multi-stage stochastic decision making problem is the end goal of this research stream.

\section{Acknowledgment}
This work has been prepared with the support of the Belgian Energy Transition Fund, project CYPRESS (\url{https://cypress-project.be/})

\bibliographystyle{IEEEtran} 
\bibliography{main}

\end{document}